\documentclass[epjST]{svjour}
\usepackage{graphics,epsfig}
\begin{document}
\title{Polymers in anisotropic environment with extended defects}
\author{V. Blavatska
\and K. Haydukivska }
\institute{Institute for Condensed Matter Physics of the National
Academy of Sciences of Ukraine, 1, Svientsitskii Str., Lviv, 79011,
Ukraine}
\abstract{ The conformational properties of flexible polymers in $d$
dimensions in environments with extended defects are analyzed both
analytically and numerically. We consider the case, when structural
defects are correlated in $\varepsilon_d$  dimensions and randomly
distributed in the remaining $d-\varepsilon_d$. Within the lattice
model of self-avoiding random walks (SAW), we apply the
pruned-enriched Rosenbluth method (PERM) and find the estimates for
scaling exponents and universal shape parameters of polymers in
environment with parallel rod-like defects ($\varepsilon_d=1$). An
analytical description of the model is developed within the des
Cloizeaux direct polymer renormalization scheme.
} 
\maketitle
\section{Introduction}
\label{intro}

Conformational properties of polymer macromolecules in solutions are
the subject of great interest in statistical polymer physics
\cite{Gennes,Grosberg,Cloizaux90}. It is established, that long
flexible polymers are characterized by a number of universal
characteristics, independent of the chemical structure and
architecture of the molecules. A typical example of such a quantity
is the averaged  end-to-end distance of $N$-monomer polymer chain
(characterizing the linear size measure of macromolecule), that
scales according to:
\begin{equation}
\langle R^2 \rangle \sim N^{2\nu}, \label{Rpure}
\end{equation}
where $\nu$ is a universal exponent, depending on space dimension
$d$ only ($\nu(d=3)=0.5877\pm0.0006$ \cite{Li95}). An other example
is the rotationally invariant characteristic of the shape of typical
polymer conformations, namely the aspherisity $\langle A_d \rangle$
\cite{Aronovitz86}, which equals $0$ for sphere-like shapes and $1$
for rod-like. For a polymer in a pure solvent in $d=3$ it was found
$\langle A_3 \rangle=0.431\pm0.002$ \cite{Yagodzinski92}.

For a long time, the  computer experiment served as  the best tool
to investigate polymer systems. In this approach, the most widely
used model was and is until now the model of self-avoiding random
walks (SAW) on a regular lattice. In spite of its simplicity, it
perfectly captures the universal conformational properties of
polymers in a solvent. Among the analytical descriptions of polymer
systems the most successful are the field-theoretical approach
\cite{Gennes}, the direct polymer renormalization \cite{Cloizaux90}
and the traditional Flory theory \cite{Gennes}. The values of
critical exponents and universal shape parameters, obtained within
the frames of these theories, are in a good agreement with results
of computer experiments.

The study of polymers in solvents in the presence of structural
impurities plays a considerable role due to the importance to
understand their behavior in colloidal solutions and near
microporous membranes \cite{Cannel80}. It also connects with the
description of proteins inside the living cells which can be treated
as disordered environments \cite{Ellis03}. The behavior of polymers
in disordered environments encountered controversies for a long
time. During more than fifteen years there has been a wide
discussion whether the presence of uncorrelated point-like defects
leads to a change in universal properties
\cite{Kremmer81,Chakrabarti81,Sahimi84,Lam84,Lyklema84,Lee88,Lee89,Kim90,Cherayil90,Roy90,Lam90,Kim91,Nakanishi91,Vanderzande92}.
This problem was first posed by Kremer \cite{Kremmer81} showing that there
exists a new value of the size exponent (\ref{Rpure}) only when
the concentration of defects is above the percolation threshold. Later,
this result was confirmed both analytically \cite{Kim83} and
numerically
\cite{Kim87,Woo91,Barat91,Grassberger93,Rintoul94,Lee96,Blavatska08}.

\begin{figure}[t!]
\begin{center}
\includegraphics{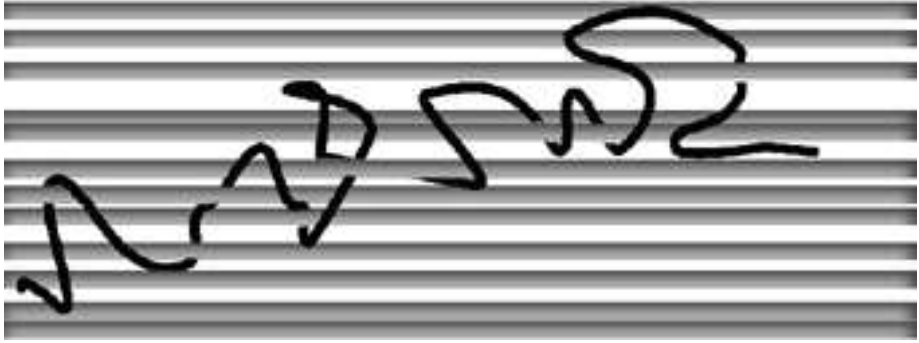}
\caption{Schematic presentation of polymer chain in an environment
with structural defects in form of parallel lines.} \label{fig:0}
\end{center}
\end{figure}

The density fluctuations of obstacles create large spatial
inhomogeneities and pore spaces, which are often of fractal
structure \cite{Dullen79}. A special case arises when fractal
extended defects of parallel orientation are present in a system,
which may cause the anisotropy of environment (see fig.
\ref{fig:0}). It is expected, that in this case there are two
scaling exponents, governing the components of the size measure
(\ref{Rpure}) in directions parallel and perpendicular to the
orientadion of the extended defects
\cite{Dorogovtsev80,Baumgaertner96}. The influence of such disorder
on the critical behavior of magnetic systems with
$\varepsilon_d$-dimensional defects of parallel orientation within
 frames of the spin $m-$vector model was analyzed in refs.
\cite{Dorogovtsev80,Boyanovsky82,Blavatska02}. It was shown that
there are two characteristic correlation lengths in the system, one
parallel and other perpendicular to the defects. As it is known
\cite{Gennes}, the  scaling properties of polymers in solution can be
described within an $m-$vector model taking the de Gennes limit (polymer
limit $m\to0$).Unfortunatly, in the anisotropic case this approach
does not give a satisfactory result.

In the present paper, we aim to obtain a quantitative description of
the scaling behavior of polymer macromolecules in environments with
$\varepsilon_d$-dimensional extended defects of parallel orientation
(causing the anisotropy of environment), applying both numerical and
analytical approaches. In the next Section, we study the special
case of defects in form of randomly placed lines of parallel orientation
($\varepsilon_d=1$), applying computer simulations. In the Section
3, we develop an analytical description of the problem within the
des Cloizeaux direct polymer renormalization scheme. We end giving
conclusions and an outlook in Section 4.

\section{Numerical analysis}
\label{NR}

\subsection{The Method}
\label{NRM}

We study the conformational properties of self-avoiding random walks
on a regular lattice with extended impurities in the form of
parallel lines (see fig. \ref{fig:0}), applying the pruned-enriched
Rosenbluth method (PERM) \cite{Grassberger97}. The method combines
the original Rosenbluth-Rosenbluth algorithm of growing chains
\cite{Rosenbluth55} and population control \cite{Wall59}. The $n$-th
monomer is placed at a randomly chosen neighbor site of the previous
$(n-1)$th monomer ($n\leq N$, where $N$ is total length of the
polymer). If this randomly chosen site is already visited by a chain
trajectory or belongs to an impurity line, it is avoided without
discarding the chain and the weight $W_n$ given to each sample
configuration at the $n$-th step is:
\begin{equation}
W_n=\prod_{l=1}^n m_l,
\end{equation}
where $m_l$ is the number of free lattice sites to place the $l$th
monomer.

 The growth is stopped when the total length $N$ of the chain is
reached (or, at $n<N$, if a the ``dead end'' without possibility to
make the next step is reached), then the next chain is started to
grow from the starting point that is chosen randomly every time.

To derive appropriate properties of the chain we apply a two-step
averaging: first average over configurations as usual and then averaging over
different realizations of disorder. The configurational average for
any quantity of interest then has the form:
\begin{eqnarray}
&&\langle (\ldots)
\rangle=\frac{1}{Z_N}{\sum_{k=1}^{M}W_N^{k}(\ldots)},
\,\,\,\,Z_N=\sum_{k=1}^{M} W_N^{{\rm conf}}, \label {R}
\end{eqnarray}
where the summation is performed over the ensemble of all
constructed $N$-step SAWs ($M\sim 10^5$ in our case). The disorder
averaging is given in the form:
\begin{eqnarray}
&&{\overline {(\ldots)}}=\frac{1}{p}{\sum_{k=1}^{p}(\ldots)},
\end{eqnarray}
where $p$ is number of replicas (we take $p=400$ in our study).

The weight fluctuations of the growing chain are suppressed in PERM
by pruning configurations with too small weights, and by enriching
samples with copies of high-weight configurations. These copies are
made while the chain is growing, and continue to grow independently
of each other. Pruning and enrichment are performed by choosing
thresholds $W_n^{<}$ and $W_n^{>}$, which are
 continuously updated as the simulation
progresses.
For updating the threshold values we apply similar rules as in
\cite{Hsu03,Bachmann03}: $W_n^{>}=C(Z_n/Z_1)(c_n/c_1)^2$ and
$W_n^{<}=0.2W_n^{>}$, where $c_n$ denotes the number of created
chains having length $n$, and the parameter $C$ controls the
pruning-enrichment statistics; it is adjusted such that on average
10 chains of total length $N$ are generated per each tour
\cite{Bachmann03}.

\subsection{Results and Discussions}
\label{NRRD}

We perform simulations for chains length up to $N=300$ monomers to measure
the end-to-end distances and up to $N=100$ monomers for the asphericity value.

Let us introduce the position vectors ${\vec r}_n=\{x_n^1,x_n^2,
x_n^3\}$ of each $n$-th monomer ($1\leq n \leq N$). In contrary to the 
the purely isotropic case (\ref{Rpure}), space anisotropy causes
different scaling behavior for the end-to-end distance components in
directions parallel ${\overline{\langle R^2_{\|} \rangle}}\equiv
{\overline{\langle (x_N^3-x_1^3)^2 \rangle}} $ and perpendicular
${\overline{\langle R^2_{\|} \rangle}}\equiv {\overline{\langle
(x_N^1-x_1^1)^2+(x_N^2-x_1^2)^2 \rangle}}$ to the lines of defects
(we assume, that the defects are extended in the $x^3$ direction). Namely:
\begin{equation}
{\overline{\langle R^2_{\|} \rangle}} \sim N^{2\nu_{\|}},\,\,\, \qquad  {\overline{\langle R^2_{\bot} \rangle}} \sim N^{2\nu_{\bot}},
\label{RR}
\end{equation}
with $\nu_{\|}\neq \nu_{\bot}$.

Analyzing our results for the end-to-end distance components the
parallel  and perpendicular to the lines of defects (see Fig. 2), we
can see that when concentration of impurity lines is small, there is
a crossover between two types of  behavior: one for short length
($\sim 40$ monomers) and the other for longer chains. We can
interpret this by the fact that at small concentration of defects
the short chain does not feel the presence of impurities (when
averaged distances between impurity lines approximately equal the
length of the chain). In such situations, to calculate the value of
size exponents we have to take into account only the second part as
it is closer to the asymptotical regime.
\begin{figure}[t!]
\begin{center}
\includegraphics{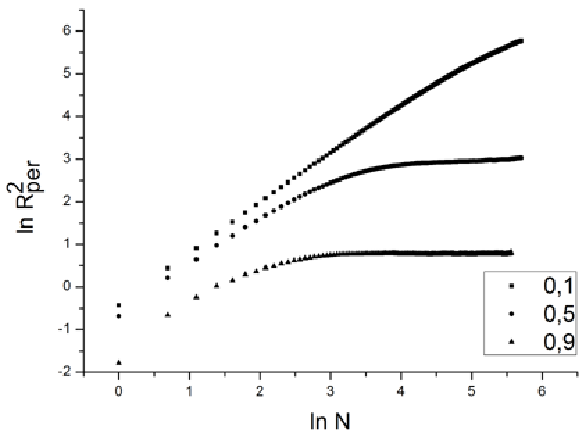}
\includegraphics{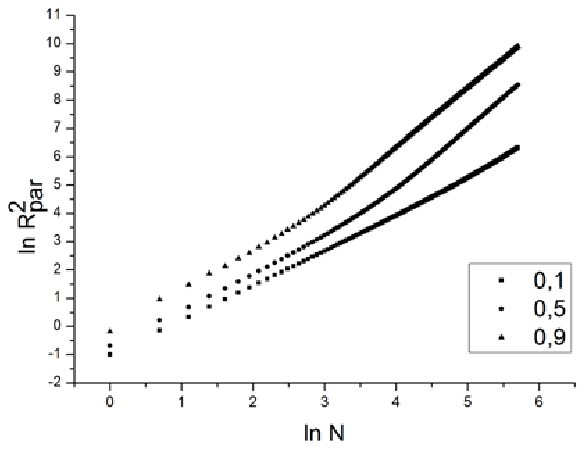}
\caption{End-to-end distance component perpendicular (left) and
parallel (right) to the impurity lines as functions of chain length
in double logarithmic scale at several values of defects
concentrations.}
\end{center} \label{fig:1}
\end{figure}

\begin{figure}[b!]
\begin{center}
\includegraphics{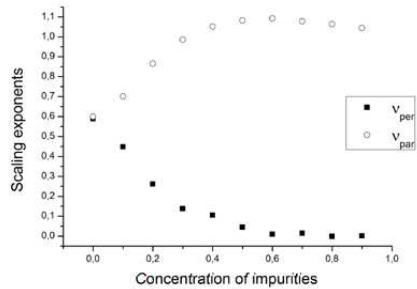}
\caption{Size exponents (\ref{RR}) at various concentration of
defects.}
\end{center}
\label{fig:3}
\end{figure}

As it was suggested, there are two different size exponents
$\nu_{\|} $ and $ \nu_{\bot}$, which can be evaluated applying the
least-square fitting  of results presented in Fig. 2 to the form
(\ref{RR}). For example, in the case of concentration 20 \% we have
$\nu_{\bot}=0.261\pm0.005$ and $\nu_{\|}=0.86\pm0.07$ which should
be compared with corresponding pure value. Results for $\nu_{\bot}$,
$\nu_{\|}$ at different concentrations of impurity lines are given
in Fig. 3. Whereas  exponent $\nu_{\|}$ gradually tends to the value
of 1 as concentration of defects grows (the polymer chain extends in
direction parallel to defects), the exponent $\nu_{\bot}$ is always
smaller and gradually tends to zero.

\begin{figure}
\includegraphics{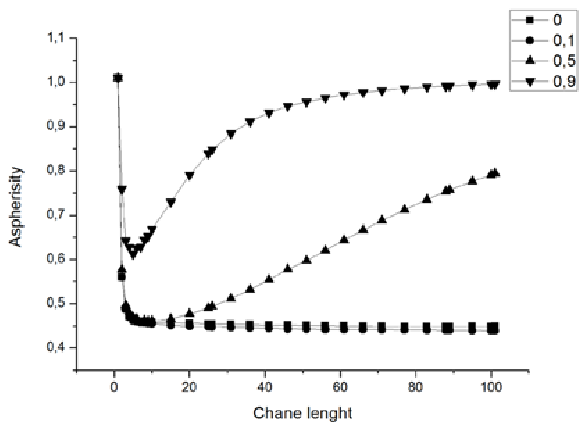}
\includegraphics{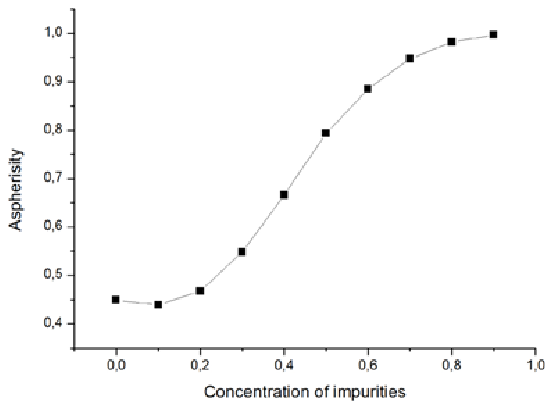}
\caption{Aspheriñity as function of the chain length at several
different concentrations of defects (left). The change of
aspheriñity with increasing the defect concentration at fixed
$N=100$ (right).} \label{fig:4}
\end{figure}

To analyze the shape of polymers we calculate the components of
gyration tensor given by:
\begin{eqnarray}
Q_{ij}=\frac{1}{N^2}\sum_{m=1}^{N}\sum_{n=1}^{N}
(x^i_n-x^i_m)(x^j_n-x^j_m)
\end{eqnarray}
where $i,j=1, 2, 3$. The shape of a typical polymer chain
conformation can be described by a quantity:
\begin{eqnarray}
A_3=\frac{3}{2}\frac{Tr\hat{Q}^{2}}{(TrQ)^{2}}\qquad
\hat{Q}=Q-\hat{I}\frac{TrQ}{3}.
\end{eqnarray}
Our results for ${\overline{\langle A_3 \rangle}}$ are given in
Figure \ref{fig:4}. One can see, that at small concentrations of
impurity lines the polymer chain does not change its shape in spite
of the existence of two different characteristic lengths. Increasing
the defect concentration to medium concentrations the shape become
more prolate and tends to a rod-like state.

As a result, polymers in an anisotropic environment with defects
aligned in a given direction are elongated in this direction. In our
case this is caused by the impurities that are extended  along this
direction and which repel the polymer.

\section{Analytical approach}

Let us start with a continuous model, where the polymer chain is
presented as a path parametrized by $\vec{r}(s)$, with $s=0\ldots
S$. An effective Hamiltonian of the system is given by:
\begin{eqnarray}
&&H=\frac{1}{2}
\int^{S}_{0}\left(\frac{d\vec{r}(s)}{ds}\right)^{2}\,ds
+u\int^{S}_{0}ds'\int^{s'}_{0}ds''\delta(\vec{r}(s')-\vec{r}(s''))\,ds+\nonumber\\
&&+\int^{S}_{0}\,V(\vec{r}(s))\,ds .
\end{eqnarray}
Note that $\vec{r}(s)$ is a $d$-dimentional vector with unit
[length] while the parameters $S$, $s$ have units [length]$^2$. $S$ is also called the Gaussian surface. 
Here, the first term describes the chain connectivity, the second term
reflects the excluded volume effect with coupling constant $u$, and
the last term arises due to the interaction between the monomers of
the polymer chain and the structural defects in the environment
given by potential $V$.

We are interested in the case, when defects are correlated in
$\varepsilon_d$ dimensions and randomly distributed in the remaining
$d-\varepsilon_d$.  The correlation function of the densities of
defects is thus assumed to be given by
\cite{Baumgaertner96,Dorogovtsev80}:
\begin{equation}
{\overline{V(\vec{r}(s))V(\vec{r}(s'))}}=v \delta^{d
-\epsilon_d}(\vec{r}(s) - \vec{r}(s')). \label{dis}
\end{equation}

Averaging the partition sum of the model over different realizations
of disorder with (\ref{dis})  results in:
 \label{AM}
\begin{eqnarray}
&&{\cal Z}_{{\rm dis}}=\int {\cal D} \{r \}\exp\left[-\frac{1}{2}
\int^{S}_{0}\left(\frac{d\vec{r}(s)}{ds}\right)^{2}\,ds
-u\int^{S}_{0}ds'\int^{s'}_{0}ds''\delta(\vec{r}(s')-\vec{r}(s''))\,ds-\right.\nonumber\\
&&\left.-v\int^{S}_{0}ds'\int^{s'}_{0}ds''\delta^{d-\epsilon_d}({r}_{d-\epsilon_d}(s')-{r}_{d-\epsilon_d}(s''))\,ds\right],
\label{zz}
\end{eqnarray}
Here, ${r}_{d-\epsilon_d}(s)$ denotes the component of position
vector $\vec{r}(s)$ perpendicular to the orientation of the extended
defects and the functional integration $\int {\cal D}\{ r \}$ is to
be performed over different chain configurations. Note that the
coupling $u$ must be positive, which corresponds to an effective
mutual repulsion of the monomers due to the excluded volume effect,
whereas the coupling $v$ should have negative sign, weakening the covolume
effect.

To evaluate quantitative estimates for the conformational
characteristics of such a model, we apply the direct polymer
renormalization scheme \cite{Cloizaux90}. All properties of interest
can be found in the form of a perturbation theory series in an
coupling constants. In particular, for the components of the
end-to-end distances (\ref{RR}) we found:
\begin{eqnarray}
&&R^2_{d-\epsilon_d}=S(d-\epsilon_d)\left(1+\frac{z_u}{(2-\frac{d}{2})(3-\frac{d}{2})}+\frac{z_v}{(2-\frac{d-\epsilon_d}{2})(3-\frac{d-\epsilon_d}{2})}\right),\\
&&R^2_{\epsilon_d}=S\epsilon_d
\left(1+\frac{z_u}{(2-\frac{d}{2})(3-\frac{d}{2})}\right),
\end{eqnarray}
here $z_u=u(2\pi)^{-d/2}S^{2-d/2}$ and
$z_v=v(2\pi)^{-(d-\varepsilon_d)/2}S^{2-(d-\varepsilon_d)/2}$ are
dimentionless couplinds. Note, that  $\varepsilon_d$ indicates the
dimentionality of the subspace parallel to the defects and
$d-\varepsilon_d$ the dimentionality of the corresponding
perpendicular subspace.

In the asymptotic limit $S\rightarrow \infty$ the physical
quantities, presented in the form of series expansions in the
coupling constants, are, however, divergent. To obtain asymptotical
values of the corresponding physical parameters a renormalization of the coupling
constants needs to be performed \cite{Cloizaux90}. The scaling
exponents attain finite values when evaluated at the stable fixed
point  of the renormalization group transformation. The fixed points
are defined as the common zeros of the flow equations, which in our
problem read:
\begin{eqnarray}
&&\beta_u=\epsilon z_u-8z_u^2-12z_uz_v,\nonumber\\
&&\beta_v=\delta z_v-8z_v^2-4z_uz_v,
\end{eqnarray}
here $\epsilon=4-d$, $\delta=\epsilon+\varepsilon_d$. We find four
fixed points:
\begin{eqnarray}
&&z^*_u=0,\quad z^*_v=0, \\
&&z^*_u=\epsilon/8,\quad z^*_v=0,\\
&&z^*_u=0,\quad z^*_v=\delta/8,\\
&&z^*_u=\epsilon/2-3\,\delta/4,\quad z^*_v=\delta/2-\epsilon/4.
\end{eqnarray}
The first of them is the well known Gaussian fixed point,
corresponding to an idealized polymer chain (random walk) without
excluded volume interactions. The second describes another well
known problem namely the polymer in a good solvent (disorder is
absent). The third point describes the Gaussian polymer in an
anisotropic environment and the last (most interesting one)
describes SAW in anisotropic environment. Note, that at
$\varepsilon_d=0$ we restore the situation with uncorrelated
point-like defects, studied previously
\cite{Kim83,Thirumalai88,Chakrabarti81,Cherayil90}. To correspond to
a physical critical point of the system, the given fixed point
should be stable and physically accessible. Unfortunately, in our
problem the two last points, which are of main interest, are
non-physical (e.g.  attain appropriate values $z_u>0$ and $z_v<0$
only in the unphysical region $\varepsilon_d<0$) and thus, cannot
provide estimates of scaling exponents. The same problem appears for
the coupling flow functions of the $m-$vector model with extended
defects \cite{Blavatska02}, when one tries to take the
$m\rightarrow0$ limit.

A similar problem of the absence of stable and physically accessible
fixed points exists also in the case of uncorrelated point-like
impurities \cite{Chakrabarti81,Cherayil90}, but it was solved by
adsorbing the interaction with disorder into the excluded volume
interaction due to special symmetry \cite{Kim83}. However it does
not work in our case of extended defects. In spite of the problems
evaluating numerical estimates of the critical exponents, our
analytical work confirms the existence of two characteristic lengths
for polymer in anisotropic environments.

\section{Conclusions}

We analyzed the conformational properties of flexible polymer chains
in disordered anisotropic environments. The anisotropy is introduced in the form of 
 $\varepsilon_d$-dimensional extended impurities of
parallel orientation. Both numerical and analytical approaches are
developed.

In numerical part of our work, we studied the scaling properties of
a SAW model on a lattice with defects in a form of lines
($\varepsilon_d=1$), extended along a fixed coordinate direction. We
conclude that there exist two critical exponents that describe the
size measure of the polymer chain in directions parallel and
perpendicular to the defects. The parallel exponent $\nu_{||}$ is
always larger than the corresponding value for polymers in a pure
solvent and gradually approaches the limit of 1 when increasing the
defect concentration. The perpendicular exponent $\nu_{\perp}$ is
always smaller and reaches the limit of 0. We also found, that
increasing the concentration of defects, the shape of the polymers
becomes more anisotropic, elongated in the direction parallel to the
extended defects and gradually reaches the limit of a rod-like
shape.

Our analytical study is developed on the basis of a continuous chain
model with applying the direct polymer renormalization scheme.
However, we encounter some controversies when analyzing stability
and physical accessibility of fixed points, corresponding to a
critical point of the system.   Clarification of this problem will
be the subject of our further work.

\end{document}